\documentclass[11pt]{article}
\usepackage[normalem]{ulem}
\usepackage{amssymb}
\usepackage{amsmath}
\usepackage{amscd}
\usepackage{latexsym}
\usepackage{color}
\usepackage{mathrsfs}

\catcode`@=11
\renewcommand{\section}{\@startsection{section}{1}{0pt}{\medskipamount}
{\medskipamount}{\large\bf}}
\numberwithin{equation}{section}
\catcode`@=12

\newcommand{\C}{\mathbb C}

\newcommand{\R}{\mathbb R}
\newcommand{\Abb}{\mathbb A}
\newcommand{\Gbb}{\mathbb G}
\newcommand{\Acal}{{\cal A}}
\newcommand{\Ecal}{{\cal E}}
\newcommand{\Fcal}{{\cal F}}
\newcommand{\Gcal}{{\cal G}}

\newcommand{\Ical}{{\cal I}}

\newcommand{\Mcal}{{\cal M}}
\newcommand{\Ncal}{{\cal N}}
\newcommand{\Tcal}{{\cal T}}
\newcommand{\Vcal}{{\cal V}}

\newcommand{\EEE}{{\mathscr{E}}}

\newcommand{\al}{\alpha}
\newcommand{\be}{\beta}
\newcommand{\ga}{\gamma}
\newcommand{\de}{\delta}
\newcommand{\eps}{\epsilon}
\newcommand{\ve}{\varepsilon}

\newcommand{\vk}{\varkappa}
\renewcommand{\th}{\theta}

\newcommand{\ph}{\phi}
\newcommand{\ch}{\chi}
\newcommand{\om}{\omega}

\def\wt{\widetilde}

\def\N2{$N{=}2$}
\def\pa{\partial}
\def\diff{\mathrm{d}}
\def\tr{\mathrm{tr}}
\def\sfrac#1#2{{\textstyle\frac{#1}{#2}}}
\def\>{\rangle}
\def\<{\langle}
\def\+{\dagger}
\def\={\ =\ }
\def\und{\quad\textrm{and}\quad}

\def\for{\quad\textrm{for}\quad}

\def\Id{\mathrm{Id}}

\newcommand{\beq}{\begin{equation}}
\newcommand{\eeq}{\end{equation}}
\newcommand{\bea}{\begin{eqnarray}}
\newcommand{\eea}{\end{eqnarray}}

\begin{document}

\begin{titlepage}
\setcounter{page}{0}


\begin{center}
{\LARGE{\bf
A low-energy limit of Yang--Mills theory\\[8pt]
on de Sitter space }}

\vspace{10mm}

\vspace{10mm}

{\Large Josh Cork,$^{\rm a}$\ Emine \c Seyma Kutluk,$^{\rm b}$\ Olaf Lechtenfeld,$^{\rm a}$\\[6pt]
and \  Alexander D. Popov$^{\rm a}$
}\\[10mm]

\noindent {\em ${}^{\rm a}$ Institut f\"ur Theoretische Physik \,{\rm and}
Riemann Center for Geometry and Physics\\
Leibniz Universit\"at Hannover, Appelstra\ss{}e 2, 30167 Hannover, Germany
}\\
\smallskip
\noindent {\em ${}^{\rm b}$ Physics Department, Middle East Technical University,\\
Dumlup{\i}nar Bulvar{\i} No:1, 06800 Ankara, Turkey }\\
\smallskip
{Email: joshua.cork@itp.uni-hannover.de, ekutluk@metu.edu.tr, olaf.lechtenfeld@itp.uni-hannover.de, alexander.popov@itp.uni-hannover.de}

\vspace{15mm}

\begin{abstract}

\noindent We consider Yang--Mills theory with a compact structure group $G$ on four-dimensional de Sitter space dS$_4$.
Using conformal invariance, we transform the theory from dS$_4$ to the finite cylinder ${\cal I}\times S^3$, where
${\cal I}=(-\pi/2, \pi/2)$ and $S^3$ is the round three-sphere. 
By considering only bundles $P\to\Ical\times S^3$ which are framed over the temporal boundary $\pa\Ical\times S^3$, we introduce additional degrees of freedom which restrict gauge transformations to be identity on $\pa\Ical\times S^3$. We study the consequences of the framing on the variation of the action, and on the Yang--Mills equations.
This allows for an infinite-dimensional moduli space of Yang--Mills vacua on dS$_4$. We show that, in the 
low-energy limit,  when momentum along ${\cal I}$ is much smaller than along $S^3$, the Yang--Mills dynamics in dS$_4$
is approximated by geodesic motion in the infinite-dimensional space $\Mcal_{\rm vac}$ of gauge-inequivalent Yang--Mills vacua on $S^3$. Since $\Mcal_{\rm vac}\cong C^\infty (S^3, G)/G$ is a group manifold, the dynamics is expected to be integrable.

\end{abstract}

\vspace{12mm}


\end{center}
\end{titlepage}

\section{Introduction}\label{sec-intro}

\noindent 
There is presently a revived and growing interest in the general allowed boundary conditions
for gauge fields at infinity, either spatial or null 
(see e.g.~\cite{Strom}--\cite{TanziGiulini} for reviews, recent developments and references). 
The question of boundary conditions is directly related to the questions of allowed solutions,
additional degrees of freedom, non-gauge symmetries of Yang--Mills theory etc..
Usually, Yang--Mills theory is considered on Minkowski space $\R^{3,1}$, which  has two null ``boundaries", past
$\cal J^-$  and future $\cal J^+$ null infinities, both with topology $\R\times S^2$. Future null infinity
$\cal J^+$ is described in the metric on $\R^{3,1}$,
\begin{equation}\label{null-inf-metric}
\diff s^2 = -\diff u^2-2\diff u\, \diff r + r^2\diff\Omega^2_2\ ,
\end{equation}
as the submanifold $r=\infty$ with the coordinates $({u}, \th , \phi )$.\footnote{Here $(r, \th ,\phi )$
are spherical coordinates in $\R^3$, $(\th ,\phi )$ are coordinates on $S^2$, $\diff\Omega^2_2$ is the metric
on the unit sphere $S^2$ and $u=t-r$.} It has boundaries ${\cal J}^+_\pm \cong S^2$ at $u=\pm\infty$. Similar to
retarded coordinates in \eqref{null-inf-metric}, one can introduce advanced coordinates with $\upsilon =t+r$ and
$\cal J^-$ as the null boundary at $r=\infty$ with coordinates $(\upsilon ,\th ,\phi )$.

For Yang--Mills theory on $\R^{3,1}$ with the structure Lie group $G$, the group of gauge transformations
$\Gcal = C^\infty (\R^{3,1}, G)$ consists of smooth maps $g:\ \R^{3,1}\to G$. Let $\Acal =\Acal_u\diff u +\Acal_r
\diff r +\Acal_\th\diff\th +\Acal_\phi\diff\phi$ be the gauge potential taking values in the Lie algebra
$\mathfrak{g}=\,$Lie$\,G$. After choosing the gauge $\Acal_u=0$ the residual group of gauge transformations
is $C^\infty (\R_+\times S^2, G)$ with $r\in \R_+$ and it contains a subgroup of maps from $S^2\subset \cal J^+$ to $G$
which may be not the identity map. It was realized that those $G$-valued functions
$g\in C^\infty (S^2, G)$, which are not unity, are not a mere redundancy of the description. The group of such
transformations (as well as the holomorphic Kac-Moody subgroup of $C^\infty (S^2, G^\C)$ for $S^2$ with punctures
where charged particles cross $S^2$) becomes a dynamical group changing states of the system. In fact, the true
group of gauge transformations is $\Gcal^0$, consisting of those transformations which become identity at the boundary. 
This boundary condition identifies the bulk principal $G$-bundle with a fixed trivial principal $G$-bundle on the boundary 
at infinity -- this identification is known as a framing. The coset space $\Gcal / \Gcal^0$ forms a group manifold of 
additional degrees of freedom that are localized at the boundary and correspond to Goldstone bosons. 
Using all these facts, Strominger and collaborators managed to reformulate soft theorems~\cite{Str,He,HMS}
for gluons and photons as Ward identities corresponding to the nontrivial asymptotic symmetries. These results
sparked interest in the study of gauge theories on manifolds with boundaries without regard to soft theorems
(see e.g.~\cite{Ser,Hen,Hos,DiP,Bal,Donn,Gom,Mat,Riello}). Our paper is also devoted to general issues of Yang--Mills theory on manifolds with boundaries, 
focused for now in the context of de Sitter space.

Our universe appears to be asymptotically de Sitter (not Minkowski) at very early and very late times. This provides
strong motivation for studying Yang--Mills theory on de Sitter space dS$_4$ -- its solutions, boundary conditions,
inequivalent ground states, infrared limit etc.. Some steps in these directions were made in \cite{Pop}--\cite{LZ}. In the aforementioned picture on Minkowski space, the dynamical symmetries act on the spatial infinities associated with the null infinities ${\cal J}^\pm$. In contrast, de Sitter space does not have a spatial infinity, and so one would expect a simple application of these established ideas to not lead to any interesting physics. Our work proposes an alternative perspective which says that this is not the end of the story for de Sitter space. Topologically, dS$_4$ is $\R\times S^3$, and it is conformally equivalent to the Lorentzian cylinder 
$(-\sfrac{\pi}{2},\sfrac{\pi}{2})\times S^3$. Hence, dS$_4$ has two temporal, spacelike conformal boundaries, both isomorphic to $S^3$. The nature of this conformal structure leads to many important and subtle differences which are not relevant in the well-studied Minkowski and asymptotically flat cases. This paper aims to address these by exploring the consequences of framing bundles over these boundaries in the context of Yang--Mills theory, and in particular to describe a low-energy limit of such a theory. We show that the classical
Yang--Mills dynamics in the infrared is described by geodesic motion in the infinite-dimensional group manifold
$C^\infty (S^3, G)/G$  of based smooth maps from $S^3\subset\,$dS$_4$ into the structure group $G$.

As a dynamical theory, one regards the Yang--Mills configuration space as a moduli space of spatial configurations \cite{BV} -- in the case of de Sitter space, this is the moduli space of connections on $S^3$ -- and dynamical trajectories are governed by tangent vectors to curves on this moduli space. The action of the group of time-independent gauge transformations splits the tangent space on the space of all spatial configurations into a ``symmetry" part, representing the true physical tangent space, and a ``gauge" part which is a physical redundancy. On the level of the Yang--Mills equations, projection onto the physical space is achieved by the Gauss law $\nabla_a{\EEE}_a=0$, where ${\EEE}_a$ is the Yang--Mills electric field. Importantly, the presence of the framing over the entire temporal boundary severely reduces the time-independent gauge group to a trivial group, which in turn changes the decomposition of the tangent space by promoting the gauge part into physical symmetries. As we shall demonstrate, in a Lagrangian approach to Yang--Mills, this necessitates an action modification by a source-like term, through which the Gauss law is relaxed in order to capture these additional degrees of freedom which otherwise would be projected away. This is complemented by an Hamiltonian picture, where these degrees of freedom extend the phase space, with the source acting as a corresponding ``momentum variable".

Pure Yang--Mills theory in four dimensions is strongly coupled in the infrared limit, and hence the perturbative expansion
for it breaks down. In the absence of a quantitative understanding of non-perturbative gauge theory, convenient alternatives
at low energy are provided by effective models among which nonlinear sigma models play an important role. In fact,
Yang--Mills, Yang--Mills--Higgs, and super-Yang--Mills theories in $d=4$ dimensions can be reduced to $d=1, d=2$ or
$d=3$ dimensions (see e.g. \cite{Man}--\cite{LP1}), depending on the choice of ``slow" and ``fast" variables in $d=4$ space-time,
by applying the adiabatic limit method. In other words, there are several low-energy limits of the same Yang--Mills theory. For 
Yang--Mills--Higgs theories (often considered as bosonic subsectors of super-Yang--Mills theories) one can also obtain 4$d$ sigma models in the strong
gauge coupling limit (see e.g. \cite{Sei}--\cite{LP2}). All these results are obtained by using an adiabatic approach and moduli space approximation.

The adiabatic approach to differential equations, based on the introduction of ``slow'' and ``fast'' variables, has existed for more than 90 years, and is used in many areas of physics. Briefly, if ``slow'' variables parametrize a space $Y$ and  ``fast'' variables
parametrize a space $Z$, then on the direct product manifold $Y\times Z$ one should introduce the metric
\begin{equation}\label{scaled-metric-eg}
g_\ve = \frac{1}{ \ve^2}\, g_Y +g_Z\ ,
\end{equation}
where $g_Y$ is a metric on $Y$,  $g_Z$ is a metric on $Z$, and $\ve\in (0, \infty )$ is a real parameter. The adiabatic limit refers
to the geometric process of scaling up a metric in some directions by sending $\ve\to 0$, whilst leaving it
fixed in the others, $g_Z$ in the case  \eqref{scaled-metric-eg}. That is, one considers the metric  \eqref{scaled-metric-eg} and the small-$\ve$  limit in
these equations.  In the simplest case $Y=\R$ with $g_Y=-1$ (time axis) one looks at solutions
to differential equations on $Z$ (``static'' solutions) and then switches on a ``slow'' dependence on time.
By using this approach, Manton has shown~\cite{Man} that, in the ``slow-motion limit'', monopole dynamics
in Minkowski space
$\R^{3,1}=\R^{0,1}\times\R^{3,0}=Y\times Z$
can be described by geodesics in the moduli space $\Mcal^n_Z$ of static $n$-monopole solutions.
In other words, it was shown~\cite{Man, AH, MS} that the Yang--Mills--Higgs model  on  $\R^{3,1}$
for slow motion reduces to a sigma model in one dimension whose target space is the $n$-monopole
moduli space $\Mcal^n_Z$ of solutions to the Yang--Mills--Higgs equations on  $Z=\R^3$.

Here we propose a framework for studying Yang--Mills theory on de Sitter space dS$_4$ which captures all additional degrees of freedom arising from a choice of framing over its conformal boundary. In doing so, this allows for a non-trivial application of the adiabatic approach to the Yang--Mills equations on de Sitter space dS$_4$. Specifically, we show that a low-energy
limit yields one-dimensional principal chiral model equations describing maps from $\R$ (time) into the group manifold
$C^\infty (S^3,G)/G$, which is identified as the moduli space of static vacua $\Mcal_{\rm vac}$ of Yang--Mills theory on dS$_4$. We argue that the
infrared dynamics is integrable.

\bigskip

\section{Yang--Mills theory on manifolds with boundary}\label{sec-YM-bdy}

\noindent {\bf Bundles and connections.}
Let $M$ be an oriented manifold  of dimension $d$, $G$ a connected and simply-connected
compact Lie group, $\mathfrak{g}$ its Lie algebra, $P$ a principal $G$-bundle over $M$, $\Acal$ a connection
one-form on $P$, and $\Fcal = \diff\Acal + \Acal\wedge\Acal$ its curvature. We also consider the bundle
of groups Int$P=P\times_G G$ ($G$ acts on itself by internal automorphisms: $h\mapsto ghg^{-1},\  h,g\in G$)
associated with $P$, and the bundle Ad$P=P\times_G\mathfrak{g}$ of Lie algebras. Both of these bundles inherit their
connection $\Acal$ from the bundle $P$. In particular, the affine connection $\nabla_\Acal$ acts (locally) on $k$-forms $\omega\in\Lambda^k(M,\mathrm{Ad}P)$ via the formula
\begin{equation}
    \nabla_\Acal\;\omega=\diff\omega+\Acal\wedge\omega+(-1)^{k+1}\omega\wedge\Acal\ .
\end{equation}

\smallskip

\noindent {\bf Symmetries.}
Let us consider the space $\Gamma (M$, Int$P)$ of global sections of the bundle Int$P$.
This space is the group of automorphisms of the bundle $P(M, G)\to M$ which induce the identity transformations of $M$.
The space $\Gamma (M$, Int$P)$ is a topological group and we consider its subgroup $\Gcal$ of smooth sections. For
trivial bundles $P=M\times G$ (direct product)  the group $\Gcal$ is
\begin{equation}\label{aut-group}
\Gcal = C^\infty (M, G)\ .
\end{equation}

We denote by $\Abb_M$ the space of connections on $P$. The infinite-dimentional group $\Gcal$ acts on $\Abb_M$ by the standard
formula
\begin{equation}\label{gt}
\Acal\ \mapsto\ \Acal^g=g^{-1}\Acal g + g^{-1}\diff g
\end{equation}
for $g\in\Gcal$. Correspondingly, the infinitesimal action of $\Gcal$ is defined by smooth global sections $\chi$ of the bundle
Ad$P$,
\begin{equation}\label{inf-gt}
\Acal\ \mapsto\ \de_\chi\Acal=\diff\chi + [\Acal ,\chi ]=: {\nabla^{}_\Acal}\,\chi
\end{equation}
with $\chi\in\,$Lie$\,\Gcal\subset\Gamma (M,$ Ad$P)$. For $P=M\times G$ we have Lie$\,\Gcal = C^\infty (M, \mathfrak{g} )$.

\smallskip

\noindent
{\bf Gauge transformations and physical symmetries.}  On a manifold $M$ with boundary $\pa M$, gauge transformations are usually considered as a subgroup
$\Gcal^0\subset\Gcal$ consisting of $G$-valued functions $g\in\Gcal^0$ which tend to the identity when approaching $\pa M$ (see e.g.~\cite{Don}).
This corresponds to a framing of the bundle $P$ over the boundary $\pa M\subset M$, namely, a fixed choice of trivialization $\varphi$ on $\pa M$, with $g\in\Gcal^0$ defined by the condition $g|_{\pa M}^\ast \varphi=\varphi$. It is easy to see that $\Gcal^0$ is a normal subgroup of $\Gcal$. The quotient group
\begin{equation}\label{bdy-gp}
\Gcal^{}_{\pa M}:=\Gcal /\Gcal^0\subset \Gamma\left(\pa M, \mathrm{Int}P|_{\pa M}\right)
\end{equation}
is easily seen, in the case of trivial bundles, to correspond to the Lie group
\begin{equation}\label{bdy-gp-trivial}
\Gcal^{}_{\pa M} =  C^\infty (\pa M, G)\ .
\end{equation}
The same logic was used recently (see e.g. \cite{Strom, Str, HMS}) for asymptotic and conformal boundaries which are not boundaries
in the strict mathematical sense. It is more accurate to talk about asymptotic conditions for fields at infinity,
but we will follow the terminology in the physics literature. To summarize, transformations \eqref{gt} on manifolds $M$
with boundaries $\pa M$ are naturally split into the gauge transformations $\Gcal^0=\{g\in\Gcal\:\mid\: g|_{\pa M}=\mathrm{Id}\}$,
and physical symmetries $\Gcal_{\pa M}$ from \eqref{bdy-gp-trivial}. The latter are sometimes called ``local" or ``large" gauge transformations, however in this paper we shall avoid this terminology, reserving any mention of the term ``gauge transformation" for the transformations in the group $\Gcal^0$, which are identity at the boundary.

\bigskip

\section{De Sitter space dS$_4$}\label{sec-dS4}

\noindent {\bf Global description.}
Four-dimensional de Sitter space can be embedded into five-di\-men\-sion\-al
Minkowski space $\R^{4,1}$ as the one-sheeted hyperboloid
\begin{equation}\label{1sheet-hyp}
\de_{ij}y^iy^j - (y^5)^2 \= R^2 \qquad\textrm{where}\quad i,j=1,\ldots,4\ .
\end{equation}
Topologically, dS$_4$ is $\R\times S^3$, that is, we can parametrize $\mathrm{dS}_4$ by $T\in\R$ and a unit four-vector $\omega\in\R^{4,0}$, specifically (see e.g. \cite{HE})
\begin{equation}\label{coordinates-ds4}
y^i= R\,\om^i\cosh T\ ,\quad y^5=R\,\sinh T
\qquad\textrm{with}\quad T\in\R
\und \de_{ij}\,\om^i\om^j=1.
\end{equation}
Our calculations do not depend on a particular choice of embedding, but for example, one might choose $\om^i=\om^i(\chi,\th,\phi)$ embedding $S^3$ into $\R^{4,0}$ by the formulae
\beq \label{embedding-S^3}
\om^1=\sin\ch\,\sin\th\,\sin\ph\ ,\quad
\om^2=\sin\ch\,\sin\th\,\cos\ph\ ,\quad
\om^3=\sin\ch\,\cos\th\ ,\quad
\om^4=\cos\ch\ ,
\eeq
where $0\le\ch,\th\le\pi$ and $0\le\ph<2\pi$.
The flat metric on $\R^{4,1}$ induces the metric on dS$_4$ in the global coordinates $(T,\chi,\theta,\phi)$ as
\beq\label{ds4-metric}
\diff s^2 \= R^2\,\bigl( -\diff T^2 + \cosh^2\! T\,\diff\Omega_3^2 \bigr)
\eeq
with
\beq\label{S^3-metric}
\diff\Omega_3^2 \=\delta_{ij}\diff\omega^i\diff\omega^j\= \diff\ch^2+\sin^2\!\ch\,(\diff\th^2+\sin^2\!\th\,\diff\ph^2)\ ,
\eeq
being the metric on the unit sphere $S^3\cong\textrm{SU}(2)$.

\smallskip

\noindent {\bf One-forms and vector fields on $S^3$.}
On $S^3$ we introduce an orthonormal basis $\{e^a\}\, ,\ a=1,2,3,$ of left-invariant one-forms satisfying
\begin{equation}\label{structure-eqn}
\diff e^a + \ve^a_{bc}\,e^b\wedge e^c\=0\ .
\end{equation}
For any choice of embedding $\{\omega^i\}$, the one-forms $\{e^a\}$ can be constructed via
\beq\label{l-i-1forms}
e^a \= -\bar\eta^a_{ij}\,\omega^i\diff\omega^j\ ,
\eeq
where $\bar\eta^a_{ij}$ is the anti-self-dual 't Hooft symbol with non-zero components
$\bar\eta^a_{bc} = \ve^{a}_{bc}$ and $\bar\eta^a_{b4} = -\bar\eta^{a}_{4b}=-\de^a_b , \ i=(a,4)$, $a=1,2,3$.
The metric on $S^3$ can then be written as
\begin{equation}\label{S^3-metric-non-hol}
\diff\Omega_3^2 =  \de_{ab} e^a e^b\ .
\end{equation}

We also introduce a basis $\{L_a\}$ of left-invariant vector fields on $TS^3$ dual to the one-forms $e^a$, which may be calculated via the corresponding formulae
\begin{equation}\label{l-i-vfs}
L_a=-\bar\eta^a_{ij}\om^i\frac{\pa}{\pa \om^j}\ , \quad L_a\lrcorner e^b=\de_a^b\ .
\end{equation}
Under commutation, these vector fields form an $\mathfrak{su}(2)$ algebra,
\begin{equation}\label{su2-algebra}
[L_a, L_b] = 2\ve^c_{ab}L_c\ .
\end{equation}
Expressions of $e^a$ and $L_a$ in terms of coordinates $(\chi , \th , \phi )$ on $S^3$ can be obtained by
substituting \eqref{embedding-S^3} into \eqref{l-i-1forms} and \eqref{l-i-vfs}.

\smallskip

\noindent {\bf Conformal coordinates.}
One can rewrite the metric \eqref{ds4-metric} on dS$_4$ in coordinates $(t,\chi , \th , \phi )=: (t,x)$ by the time reparametrization~\cite{HE}
\begin{equation}\label{dS4-conf-change}
t\=\arctan (\sinh T)\=2\arctan(\tanh\sfrac{T}{2})
\qquad\Longleftrightarrow\qquad
\frac{\diff T}{\diff t} \= \cosh T \= \frac{1}{\cos t}\ ,
\end{equation}
in which $T\in (-\infty, \infty )$ corresponds to $t\in (-\sfrac{\pi}{2}, \sfrac{\pi}{2})$.
The metric \eqref{ds4-metric} in these coordinates reads
\begin{equation}\label{dS4-metric-conformal}
\diff s^2 \= \frac{R^2}{\cos^2\!t}\,\bigl(- \diff t^2 + \de_{ab}e^ae^b \bigr)
\= \frac{R^2}{\cos^2\!t}\,\diff s^2_{\textrm{cyl}}  \ ,
\end{equation}
where
\begin{equation}\label{cylinder-metric}
\diff s^2_{\textrm{cyl}} \= - \diff t^2 + \diff\Omega^2_3
\end{equation}
is the standard metric on the Lorentzian cylinder $\Ical\times S^3$, where $\Ical$ is the interval
$(-\sfrac{\pi}{2}, \sfrac{\pi}{2})$ parametrized by $t$.

\smallskip

\noindent
{\bf Boundary of dS$_4$.} The conformal boundary of dS$_4$ consists of the two 3-spheres at $t=\pm\frac{\pi}{2}$
or, equivalently, at $T=\pm\infty$. This is the true boundary of the cylinder $M=(-\sfrac{\pi}{2},\sfrac{\pi}{2})\times S^3$,
\begin{equation}\label{bdy-cylinder}
\pa ((-\sfrac{\pi}{2},\sfrac{\pi}{2})\times S^3)=S^3_{t={+}\frac{\pi}{2}}\sqcup S^3_{t={-}\frac{\pi}{2}}\ .
\end{equation}

\bigskip

\section{Gauge theory on $\Ical\times S^3$}\label{sec-GT-on-M}
\noindent {\bf Conformal invariance.} We are ultimately concerned with studying the Yang--Mills equations on de Sitter space. However, because in four dimensions the Yang--Mills equations and action are conformally
invariant, their solutions on de Sitter space can be obtained by solving
the equations on $\Ical\times S^3$ with the cylindrical metric \eqref{cylinder-metric}. Thus, henceforth our objects of interest are defined over $\Ical\times S^3$.

\smallskip
\noindent {\bf Connections and curvature.} Let $P(M,G)$ be a principal $G$-bundle over $M=\Ical\times S^3$. We assume $G$ is compact and simply connected, which ensures $\pi_2(G)=0$ \cite{Stern}, and since $\Ical$ is contractible, this means $P$ is trivial. Let $\Acal$ be a gauge potential for a connection on $P$, which we can express in the basis $(e^\mu)$ for the cotangent bundle $T^\ast M$, with $e^0=\diff t$, and $e^a$ given in \eqref{l-i-1forms}, as 
\begin{equation}\label{connection-1form}
    \Acal=\Acal_t\;\diff t+\Acal_a\;e^a\ .
\end{equation}
In this basis, in contrast to the situation in a coordinate basis, the curvature $2$-form $\Fcal=\diff\Acal+\Acal\wedge\Acal$ picks up an additional term according to the structure equations \eqref{structure-eqn}. Specifically, writing
\begin{align}
    \Fcal&=\dfrac{1}{2}\Fcal_{\mu\nu}\;e^\mu\wedge e^\nu= \Fcal_{ta}\;\diff t\wedge e^a+\dfrac{1}{2}\Fcal_{ab}\;e^a\wedge e^b\label{curvature-general}
\end{align}
and introducing the covariant derivatives
\begin{equation}\label{covariant-derivatives}
    \nabla_t=\pa_t+[\Acal_t,\cdot], \und \nabla_a=L_a+[\Acal_a,\cdot]\ ,
\end{equation}
the components of $\Fcal$ in \eqref{curvature-general} can be understood as
\begin{align}
    \Fcal_{ta}&=[\nabla_t,\nabla_a]=\pa_t\Acal_a-\nabla_a\Acal_t\ ,\label{curvature-1}\\
    \Fcal_{ab}&=[\nabla_a,\nabla_b]-2\ve_{ab}^c\nabla_c=L_a\Acal_b-L_b\Acal_a+[\Acal_a,\Acal_b]-2\ve^c_{ab}\Acal_c\ ,\label{curvature-2}
\end{align}
where $L_a$ are the left-invariant vector fields \eqref{l-i-vfs} dual to the $e^a$.
\smallskip

\noindent {\bf Gauge transformations and symmetries.} We denote by $\Abb_M$ the space
of connections on $P$ and by $\Gcal$ the subgroup of smooth automorphisms $\Gamma(M,\mathrm{Int}P)$ of the bundle $P$ acting on $\Abb_M$
by formula \eqref{gt}. By triviality of $P$, we have
\begin{equation}\label{aut-group-cyl}
\Gcal = C^\infty (\Ical\times S^3,G)\ .
\end{equation}
In Section \ref{sec-dS4} we introduced coordinates $(t,x)= (t, \chi , \th , \phi )$
on $\Ical\times S^3$, and any $g\in\Gcal$ may be viewed as a $G$-valued function of these coordinates. Since $M=\Ical\times S^3$ is a manifold with boundary, specifically two components both isomorphic to $S^3$, as discussed in Section \ref{sec-YM-bdy}, it is natural to frame $P$ over the boundary, which means choosing a fixed isomorphism $P|_{\pa M}\to(S_+^3\sqcup S_-^3)\times G$. The gauge group $\Gcal^0$ is the subgroup of $\Gcal$ which fixes the framing, and this is isomorphic to
\begin{equation}\label{gauge-group}
\Gcal^0 = \bigl\{g\in\Gcal\ \mid \ g(t=\pm\,\sfrac{\pi}{2})=\Id\bigr\}\ .
\end{equation}
We consider two connections equivalent if and only if they differ via \eqref{gt} by an element $g\in\Gcal^0$. The gauge group $\Gcal^0$ is a normal subgroup of $\Gcal$, and the quotient group is
\begin{equation}\label{bdy-transformations}
\Gcal/\Gcal^0:=\Gcal^{}_{\pa M} \cong C^\infty (S^3, G)\times C^\infty (S^3, G)=: \Gcal^{+}_{\pa M}\times \Gcal^{-}_{\pa M}\ ,
\end{equation}
as follows from the discussion in Section \ref{sec-YM-bdy}. The transformation \eqref{gt} is well-defined for elements $g\in\Gcal_{\pa M}$, with the two groups $\Gcal_{\pa M}^\pm$ acting independently on the two disjoint copies of $S^3$  at the conformal past and future spacelike infinities; this is not so in the Minkowski case because $S^2\cong{\cal J}^-_+$ and $S^2\cong{\cal J}^+_-$ intersect at a point, namely spatial infinity $i^0$, and so the physical symmetries must be non-trivially matched at this point (see e.g. \cite{Strom} and references within). Furthermore, in general the action of $\Gcal_{\pa M}$ is non-trivial on the moduli space $\Abb_M/\Gcal^0$. Indeed, from \eqref{gt}, $[\Acal^g]=[\Acal]$ for $g\in\Gcal_{\pa M}$ if and only if $g$ is (covariantly) constant with respect to $\Acal|_{\pa M}$. Since this requires $g$ to have constant eigenvalues on $\pa M$, this of course means the fixed-point set is trivial. For this reason, $\Gcal_{\pa M}$ is a considered as a dynamical symmetry group acting on the set of connections $\Abb_M$.

\smallskip

\noindent {\bf Gauge choices and holonomy.} One can always relate the time-like component $\Acal_t$ of the connection $\Acal$ to one such that $\Acal_t=0$ by transforming $\Acal$ via \eqref{gt} with the symmetry group \eqref{aut-group-cyl}. For
that one has to solve the parallel transport equation
\begin{equation}\label{pt-eqn}
\pa_t h + \Acal_t h =0
\end{equation}
with the boundary condition
\begin{equation}\label{pt-bdy-cond}
h(t=-\sfrac{\pi}{2}\ ,\ x) = \Id\ .
\end{equation}
The condition \eqref{pt-bdy-cond} is generic since solutions $h(t,x)$ to \eqref{pt-eqn} are invariant under right-multiplication by any $k\in \Gcal$ with $\pa_tk=0$, and so $h(t,x)$ and $h(t,x)h(t=-\sfrac{\pi}{2},x)^{-1}$ define the same component $\Acal_t$ of $\Acal$. For any given
$\Acal_t$, the unique solution of the equation \eqref{pt-eqn} with the boundary condition \eqref{pt-bdy-cond} is given by
\begin{equation}\label{pt-soln}
h(t, x) = {\cal P}\exp \Bigl(-\int^t_{-\sfrac{\pi}{2}}\Acal_{t'}(t',x)\;\diff t'\Bigr)\ ,
\end{equation}
where $\cal P$ denotes path ordering. The group element
\begin{equation}\label{holonomy}
\Omega(x)=h(t=\sfrac{\pi}{2}, x) = {\cal P}\exp \Bigl(-\int^{\sfrac{\pi}{2}}_{-\sfrac{\pi}{2}}\Acal_{t'}(t',x)\;\diff t'\Bigr)\ ,
\end{equation}
is the holonomy of $\Acal$ along $\Ical$ in the space $\Ical\times S^3$.

The framing of the bundle $P$ removes gauge freedom at $t=\pm\sfrac{\pi}{2}$, and this is manifested here in the holonomy \eqref{holonomy}. This is analogous to the situation when one considers Yang--Mills theory on a circle \cite{HeHo}, where the holonomy around the circle (the Wilson line) is a physical degree of freedom. A connection $\Acal\in\Abb_{M}$ is said to have trivial holonomy if $\Omega(x)=\mathrm{Id}$ for all $x\in S^3$. This is equivalent to the condition that $h\in\Gcal^0$, and only in this situation may the gauge choice $\Acal_t=0$ be made. In general, for connections with non-trivial holonomy, we are forced to relax this choice to $\pa_t\Acal_t=0$. To see this, let $h$ solve \eqref{pt-eqn} with \eqref{pt-bdy-cond}, and let $\Theta\in C^\infty(S^3,\mathfrak{g})$ be such that $\Omega(x)=h(t=\sfrac{\pi}{2},x)=\exp(-\pi\Theta(x))$. Then $g\in\Gcal^0$ given by
\begin{equation}\label{gt-fixing-A_t}
    g(t,x)=h(t,x)\exp\left(\left(t+\dfrac{\pi}{2}\right)\Theta(x)\right)
\end{equation}
sets $\Acal^g$ so that $\Acal_t^g=\Theta$, which is time-independent.

In this paper we shall consider only the trivial holonomy case, with the more general problem of non-trivial holonomy reserved for a later work. With this simplification, the time-like component is always of the form $\Acal_t=g_0^{-1}\pa_tg_0$ for some $g_0\in\Gcal^0$, and thus, as discussed above, we can choose the gauge where $\Acal_t=0$. Restricting to the class of connections with trivial holonomy also breaks the group of physical symmetries \eqref{bdy-transformations} to its diagonal subgroup, identified as \begin{equation}\label{GS3}
    \Gcal_{S^3}:=C^\infty(S^3,G)\ .
\end{equation}

\smallskip

\noindent {\bf The dynamical configuration space.} The dynamics of Yang--Mills theory on $M=\Ical\times S^3$ are governed by ``paths" in a moduli space of connections on a corresponding principal bundle $P_{S^3}$ over $S^3$, which for the same reason as above, must be trivial. The imposition of the framing over the boundary of $M$ forces modifications to Yang--Mills theory which we introduce later in Section \ref{sec-YM-on-M}. To motivate these modifications, it is worth reviewing the geometry of this Yang--Mills configuration space \cite{BV}, and how the framing affects it.

Let $\Abb_{S^3}$ denote the space of all connections on $P_{S^3}$. This is an affine space over $\Lambda^1(S^3$, Ad$P_{S^3})$, hence for each $\Acal \in\Abb_{S^3}$, we have a canonical identification between the
tangent space $T_\Acal \Abb_{S^3}$ and the space $\Lambda^1(S^3$, Ad$P_{S^3}$) of one-forms on $S^3$  with values in the
Lie algebra $\mathfrak{g}$ of the Lie group $G$. The group $\Gcal_{S^3}$ acts on $\Abb_{S^3}$ via \eqref{gt}, and vectors which are tangent to the full $\Gcal_{S^3}$ action lie in the vertical space
\begin{equation}
    V_\Acal\Abb_{S^3}:=\mathrm{im}(\nabla_{\Acal})\subset\Lambda^1(S^3,\mathrm{Ad}P_{S^3})\ ,
\end{equation}
consisting of the ``infinitesimal gauge transformations" spanned by one-forms of the form \eqref{inf-gt}. The tangent space to the quotient space $\Abb_{S^3}/\Gcal_{S^3}$ is identified as the space of horizontal vectors complementary to $V_\Acal\Abb_{S^3}$. This is determined by a choice of \textit{connection} on the ``$\Gcal_{S^3}$-bundle" $\Abb_{S^3}$. There is a canonical choice in the presence of an inner-product on $T_\Acal\Abb_{S^3}$. Since we consider matrix groups $G$, the metric on $\mathfrak{g}$
is defined by the trace
$\tr$, and so the metric on $S^3$ and on $\mathfrak{g}$ induce a natural inner product on $T_\Acal\Abb_{S^3}=\Lambda^1(S^3$, Ad$P_{S^3})$, namely the $L^2$-inner-product defined by
\begin{equation}\label{metric-1forms-S^3}
\langle \xi_1, \xi_2\rangle_{L^2}=-\int_{S^3}\diff V_3\
\tr(\xi_{1a}\, \xi_{2a})\ ,
\end{equation}
for $\xi_1=\xi_{1a}e^a ,\ \xi_2=\xi_{2a}e^a \in\Lambda^1(S^3, \mathfrak{g} )$. With this, the horizontal space may be identified as
\begin{equation}
    H_\Acal\Abb_{S^3}:=\ker(\nabla_{\Acal}^\ast)\subset\Lambda^1(S^3,\mathrm{Ad}P_{S^3})\ ,
\end{equation}
where $\nabla_{\Acal}^\ast:\Lambda^1(S^3$, Ad$P_{S^3})\to\mathrm{Lie}(\Gcal_{S^3})$ is the $L^2$-adjoint to $\nabla_{\Acal}$, i.e. such that
\begin{align}
    \langle\nabla_{\Acal}^\ast\xi,\chi\rangle_{L^2}=\langle\xi,\nabla_{\Acal}\chi\rangle_{L^2}\ ,
\end{align}
for all $\xi\in\Lambda^1(S^3,\mathfrak{g})$ and $\chi\in\mathrm{Lie}(\Gcal_{S^3})$. In this way, the decomposition $T_\Acal\Abb_{S^3}=V_\Acal\Abb_{S^3}\oplus H_\Acal\Abb_{S^3}$ is $L^2$-orthogonal. In components, the horizontal vectors are determined by solutions $\xi_a$ to the equation
\begin{equation}\label{horizontal-proj}
    \nabla_a\xi_a=0\ .
\end{equation}
The equation \eqref{horizontal-proj} appears analogously in the Yang--Mills equations as the Gauss law
\begin{equation}\label{Gauss-law}
    \nabla_a\Fcal_{at}=0\ ,
\end{equation}
which is one of the Euler-Lagrange equations for the pure Yang--Mills action. After the gauge choice $\Acal_t=0$ is made (which we can do since we assume trivial holonomy), for each $t\in\Ical$, $\EEE_a:=\Fcal_{ta}=\pa_t\Acal_a$ are seen as the components for a $\mathfrak{g}$-valued one-form on $S^3$ representing a tangent vector to a curve $t\mapsto\Acal_a(t)$ in $\Abb_{S^3}$, that is, a dynamical trajectory. The Gauss law \eqref{Gauss-law} is not a dynamical equation, but rather a constraint, which by the above description plays the role of projecting the dynamics from the full configuration space $\Abb_{S^3}$ to the moduli space $\Abb_{S^3}/\Gcal_{S^3}$.

This fact forces a reassessment of the Yang--Mills equations when one considers the role of the framing; due to the framing over the boundary of $M$, the group $\Gcal_{S^3}$ is not a group of gauge transformations, but a physical symmetry group, and so the true physical configuration space in this scenario should be all of $\Abb_{S^3}$, with configurations which differ by an element of $\Gcal_{S^3}$ considered as physically distinct. Coupling this with the above picture, the upshot is, in order to account for the additional $\Gcal_{S^3}$ degrees of freedom imposed by the framing, one is forced to modify the Yang--Mills action so that the Gauss law \eqref{Gauss-law} is not one of the equations of motion. This we discuss in the following section.

\bigskip

\section{Yang--Mills theory on the framed bundle $P\to\Ical\times S^3$}\label{sec-YM-on-M}

\noindent {\bf Action functional.}
In the non-holonomic basis described by $\diff t=e^0$ and \eqref{l-i-1forms}, the pure Yang--Mills action on $M=\Ical\times S^3$ is
\begin{align}
    S_{\rm YM}(\Acal)&:=\dfrac{1}{2e^2}\int_M\tr(\Fcal\wedge\star_M\Fcal)\notag\\
    &=-\dfrac{1}{4e^2}\int_{\Ical\times S^3}\diff t\wedge \diff V_3\;\tr(2\Fcal_{ta}\Fcal_{ta}-\Fcal_{ab}\Fcal_{ab})\ ,\label{YM-action}
\end{align}
with $e>0$ the gauge coupling constant, $\diff V_3:=e^1\wedge e^2\wedge e^3$ the volume form on $S^3$, and the components of the curvature given by \eqref{curvature-1}-\eqref{curvature-2}. To account for the additional degrees of freedom introduced by the framing of $P$, as motivated in the previous section, we modify the Yang--Mills action with the addition of an external static source $j=j_t\;\diff t\in\Lambda^1(M,\mathfrak{g})$. We hence consider the action $S(\Acal)=S_{\rm YM}(\Acal)+S_j(\Acal)$ where
\begin{align}
    S_j(\Acal):=\frac{1}{e^2}\int_M\tr(\Acal\wedge\star_M j)=-\frac{1}{e^2}\int_{\Ical\times S^3}\diff t\wedge \diff V_3\;\tr(\Acal_tj_t)\ .\label{source-term-action}
\end{align}
The source transforms with respect to gauge transformations as
\begin{equation}
    j\ \mapsto\ j^g:=g^{-1}jg\ ,\label{gt-source}
\end{equation}
and so in order to preserve $\Gcal^0$-invariance of the action, we must impose that $j$ is covariantly constant. Since $j$ is static, i.e. $j^a=0$, this is realized by the condition
\begin{equation}
    \nabla_tj_t=0\ .\label{cov-const-source}
\end{equation}
Explicitly, \eqref{cov-const-source} is solved by
\begin{align}
    j_t(t,x)=h(t,x)\lambda(x) h(t,x)^{-1},\quad\lambda(x)=j_t(-\sfrac{\pi}{2},x)\ ,\label{jt}
\end{align}
where $h$ is the parallel transport operator \eqref{pt-soln}. Since we assume trivial holonomy, we may always choose a gauge where $j_t=\lambda$, and this choice is equivalent to fixing $\Acal_t=0$. In general, the source $j$ is included in the action to encode the additional physical degrees of freedom arising as a result of the framing, namely the holonomy \eqref{holonomy} at $t=\sfrac{\pi}{2}$, and $\lambda\in C^\infty(S^3,\mathfrak{g})$, a fixed external field injected by the framing of the bundle $P$ at $t=-\sfrac{\pi}{2}$.

\smallskip

\noindent {\bf The moduli space of Yang--Mills connections.} The Euler-Lagrange equations for the action $S=S_{\rm YM}+S_j$ may be written concisely as $\nabla_\Acal\star_M\Fcal+\star_M j=0$. In components, this reads as the dynamical equations
\begin{equation}
    \nabla_t\Fcal_{at}=\nabla_b\Fcal_{ab}+\ve_{bc}^a\Fcal_{bc}\ ,\label{YM-eqn-1}
\end{equation}
and the relaxed Gauss law
\begin{equation}
    \nabla_a\Fcal_{at}=j_t\ .\label{relaxed-Gauss-law}
\end{equation}
Here, as with the curvature \eqref{curvature-1}, the additional term in \eqref{YM-eqn-1} arises due to the choice of non-holonomic basis. The holonomy $\Omega(x)$ and the external field $\lambda(x)$ are fixed $\Gcal^0$-invariant quantities, and for each choice we have a distinct moduli space of Yang--Mills connections on $M=\Ical\times S^3$, which we denote by $\Mcal_{\rm YM}(\Omega,\lambda)$. In this paper we consider only the special case $\Mcal_{\rm YM}(\Id,\lambda)$. These are defined by the solutions to the Yang--Mills equations, up to equivalence by the action \eqref{gt} of $\Gcal^0$. 

\smallskip

\noindent {\bf Generalized variations and restricted Yang--Mills equations.}  The equations \eqref{YM-eqn-1}-\eqref{relaxed-Gauss-law} arise from the variation of the action with respect to variations $\de\Acal$ which vanish on the boundary $\pa M$. Since $S^3$ has no boundary, this assumption need not be made for the variations $\de\Acal_t$ with respect to $\Acal_t$, however, general variation with respect to the field $\Acal_a$ leads to a boundary term proportional to
\begin{equation}\label{var-YM-action}
\de S_{\pa M}\sim\left.\int_{S^3}\diff V_3\
\tr(\de\Acal_{a}\, \Fcal_{at})\right|^{\sfrac{\pi}{2}}_{-\sfrac{\pi}{2}}\ .
\end{equation}
If we allow for variations which have the form 
\begin{equation}\label{allowed-variations}
\de_\chi\Acal_a=\nabla_a\chi\for \chi\in C^\infty(S^3, {\mathfrak{g}})\ ,
\end{equation}
which corresponds to an infinitesimal action \eqref{inf-gt} of the subgroup $\Gcal_{S^3}$, defined by \eqref{GS3}, of the dynamical symmetry group \eqref{bdy-transformations}, the integral
\eqref{var-YM-action} over $S^3$ for variations \eqref{allowed-variations} can be written as
\begin{equation}
    \int_{S^3}\diff V_3\
    \tr(\nabla_a\chi\, \Fcal_{at})=-\int_{S^3}\diff V_3\
    \tr(\chi\,j_t)\ ,
\end{equation}
where we have used equation \eqref{relaxed-Gauss-law}, and the fact that $\pa S^3=\emptyset$. Since we assume trivial holonomy, \eqref{var-YM-action} will vanish, which can be seen from the formula \eqref{jt}. So the dynamical equations \eqref{YM-eqn-1} arise even for variations of the form \eqref{allowed-variations}. On the other hand, the relaxed Gauss law \eqref{relaxed-Gauss-law} is the variational equation for $\Acal_t$, assuming variations which vanish identically on $\pa M$. However since $\Acal_t=g_0^{-1}\pa_tg_0$, for some $g_0\in\Gcal^0$, then the allowed variations may be of the form
\begin{equation}
    \de_\eps\Acal_t=\nabla_t\eps\for\eps\in\mathrm{Lie}(\Gcal^0)\ .\label{variations-t-trivial-holonomy}
\end{equation}
In this case, if we assume \eqref{YM-eqn-1} holds, then the variation of the action $S=S_{\rm YM}+S_j$ with respect to \eqref{variations-t-trivial-holonomy} vanishes trivially since $\nabla_t(\nabla_a\Fcal_{at}-j_t)=0$ due to \eqref{cov-const-source} and the identity $[\Fcal_{\mu\nu},\Fcal_{\mu\nu}]=0$, coupled with the definitions \eqref{curvature-1}-\eqref{curvature-2}. Therefore in the trivial holonomy case, and where we only allow variations of the form \eqref{variations-t-trivial-holonomy}, the equation \eqref{relaxed-Gauss-law} is not an equation of motion, and should not be considered as part of the Yang--Mills equations. To see this from another perspective, one can view the term $S_j$ in the action as a gauge-fixing term for the gauge choice $\Acal_t=0$, with the field $\lambda$ a Lagrange multiplier for this constraint. The argument above says that this field can be arbitrary if we only allow variations of the form \eqref{variations-t-trivial-holonomy}. Alternatively, we can view the equation \eqref{relaxed-Gauss-law} as the definition of $\lambda$, after the gauge choice $\Acal_t=0$ is imposed. Either way, we see that the source-like term $S_j$ removes the strict Gauss law \eqref{Gauss-law} from the Yang--Mills equations, and the issue of projecting out the $\Gcal_{S^3}$ degrees of freedom is avoided.

On the other hand, the relaxed Gauss law \eqref{relaxed-Gauss-law} also allows us to determine $\Acal_t$ in terms of $\Acal_a$ and $j_t$. Indeed, we have
\begin{equation}\nonumber
\Fcal_{at} = \nabla_a\Acal_t-\pa_t\Acal_a\quad\Rightarrow\quad\nabla_a\Fcal_{at}=
\nabla^2\Acal_t - \nabla_a\pa_t\Acal_a=j_t
\end{equation}
\begin{equation}\label{resolving-Gauss-law}
\Leftrightarrow\quad\nabla^2\Acal_t =j_t+ \nabla_a\pa_t\Acal_a\ ,
\end{equation}
where $\nabla^2=\nabla_a\nabla_a$. Thus, given fixed $\Acal_a$ and $j_t$, $\Acal_t$ is determined by resolving the covariant Poisson equation \eqref{resolving-Gauss-law} on $S^3$.

\smallskip

\noindent {\bf The Hamiltonian picture.} To further appreciate the role of the modifications described above, it is useful to briefly discuss how these fit into an Hamiltonian formulation of Yang--Mills theory \cite{FadSlav}. Recall that in this picture, after imposing the gauge $\Acal_t=0$, one introduces the conjugate variables $(\Acal_a,\EEE_a)$ defined on $S^3$, with $\EEE_a=\Fcal_{ta}=\pa_t\Acal_a$, which act as coordinates for an infinite-dimensional Poisson manifold $\Mcal$, with Poisson bracket
\begin{equation}
    \left\{\EEE_a^I(x),\Acal_b^J(x)\right\}=\delta_{ab}\delta^{IJ}\delta(x-y)\ ,
\end{equation}
where we have written $\EEE=\EEE_a^IX_Ie^a$ and $\Acal=\Acal_a^IX_Ie^a$ in terms of a basis $X_I$ for $\mathfrak{g}$. By introducing the framing, we are extending this phase space to
\begin{equation}\label{ex-phase-space}
\wt{\Mcal}=\{(\Acal_a,\EEE_a,g,\lambda)\}\ ,
\end{equation}
with the additional conjugate pair $(g,\lambda)\in\Gcal_{S^3}\times C^\infty(S^3,\mathfrak{g})$ forming coordinates for the extension: the cotangent bundle $T^\ast\Gcal_{S^3}$.

\smallskip

\noindent {\bf Noether charges, symmetries, and the Gauss law.} For any Cauchy surface $\Sigma$ (in our case, $\Sigma=S^3$), the conserved Noether charges in Yang--Mills theory are defined for all $\chi\in C^\infty(\Sigma,\mathfrak{g})$ as
\begin{equation}\label{noether-charges-general}
    Q_\Sigma(\chi):=\int_{\pa\Sigma}\tr(\chi \EEE_{\pa\Sigma})-\int_\Sigma\diff{\rm Vol}_\Sigma\, \tr(\chi C)\ ,
\end{equation}
where $\EEE_{\pa\Sigma}$ is the component of the electric field orthogonal to the boundary $\pa\Sigma$, and
\begin{equation}\label{Gauss-constraint}
    C(\Acal,\EEE)=\nabla^\ast_\Acal\EEE\ .
\end{equation}
In our case $\Sigma=S^3$, so $\pa\Sigma=\emptyset$ and the charges \eqref{noether-charges-general} reduce to
\begin{equation}\label{noether-charges-S^3}
    Q_{S^3}(\chi):=-\int_{S^3}\diff{\rm V}_3\, \tr(\chi C)\ ,
\end{equation}
for $\chi\in\mathrm{Lie}(\Gcal_{S^3})$, and $C=\nabla_a\EEE_a$. In the Hamiltonian picture introduced above, off-shell, these charges generate the infinitesimal action of $\Gcal_{S^3}$ on the ordinary phase space $\Mcal$ via
\begin{equation}
    \{Q_{S^3}(\chi),\EEE_a\}=[\chi,\EEE_a],\quad\{Q_{S^3}(\chi),\Acal_a\}=\nabla_a\chi\ .
\end{equation}
From the perspective of symplectic geometry, the function $Q_{S^3}=C^\flat:\Mcal\to\mathrm{Lie}(\Gcal_{S^3})^\ast$ is a moment map for the Hamiltonian action of $\Gcal_{S^3}$ on $\Mcal$. The ordinary Gauss law \eqref{Gauss-law} is defined by $C=0$, and analogously to the discussion at the end of Section \ref{sec-GT-on-M}, usually one considers the submanifold $C^{-1}(0)\subset\Mcal$, and the physical space as the symplectic reduction
\begin{equation}
    \Mcal_{\rm phys}=\Mcal//\Gcal_{S^3}=C^{-1}(0)/\Gcal_{S^3}\ ,
\end{equation}
on which the charges \eqref{noether-charges-S^3} would vanish. In contrast, we consider the extended phase space $\wt{\Mcal}$ as in \eqref{ex-phase-space}. There is still an action of $\Gcal_{S^3}$ on $\wt{\Mcal}$ related to the charges $Q_{S^3}(\chi)$, which are non-zero as a result of the relaxed Gauss law \eqref{relaxed-Gauss-law}. In particular, the physical space is thus modified by considering instead
\begin{equation}
    \wt{C}=C+\lambda,\quad\wt{C}^{-1}(0)\subset\wt{\Mcal}\ .
\end{equation}
Furthermore, since $\lambda$ may be defined by \eqref{relaxed-Gauss-law} in terms of $(\Acal_a,\EEE_a)$, it is not independent, and so we may always reach the ordinary phase space $\Mcal$ by acting with $\Gcal_{S^3}$ in such a way which sets $g=1$ in \eqref{ex-phase-space}.

\smallskip

\noindent {\bf Newtonian mechanics form of the action.} The pure Yang--Mills action \eqref{YM-action} can be rewritten as
\begin{equation}\label{YM-action_T-V}
S_{\rm YM}(\Acal) \= \frac{1}{e^2} \int_{{\cal I}}\diff t\; \bigl(\Tcal(\Acal)-\Vcal(\Acal)\bigr)\ ,
\end{equation}
where
\begin{equation}\label{kinetic-energy}
\Tcal(\Acal) = - \sfrac12\int_{S^3}\diff V_3\
\tr(\Fcal_{ta}\Fcal_{ta})\ ,
\end{equation}
and
\begin{equation}\label{potential-energy}
\Vcal(\Acal) = - \sfrac14\int_{S^3}\diff V_3\
\tr(\Fcal_{ab}\Fcal_{ab})\ .
\end{equation}
are the kinetic and potential energies respectively. As detailed at the end of Section \ref{sec-GT-on-M}, a solution of the full Yang--Mills equations on
 $\Ical\times S^3$ (a dynamic solution) can be considered as a smooth path $\Acal : \ t\mapsto\Acal(t)$ in an infinite-dimensional configuration space, whose motion is governed by the Yang--Mills equations. 
 After resolving the constraint specified by the relaxed Gauss law \eqref{relaxed-Gauss-law} via \eqref{resolving-Gauss-law}, the non-dynamical variable $\Acal_t$ is determined, and the action \eqref{YM-action_T-V} may be considered as the classical action for a
``particle" $\Acal(t)=\Acal_a(t)\;e^a$ with kinetic energy \eqref{kinetic-energy} and potential energy \eqref{potential-energy} (cf. \cite{BV,Man}).
\smallskip

\noindent {\bf Slow motion.} We wish to study the slow motion of $\Acal(t)$. To do this, we consider the so-called ``slow time" \cite{Serg,Stu, Uhl}:
\begin{equation}\label{slow-time}
\tau : = \ve\, t\ ,
\end{equation}
where $\ve >0$ is a small parameter. This rescales the interval $\Ical$ to $\Ical_\ve=(-\sfrac{\ve\pi}{2},\sfrac{\ve\pi}{2})$, and the action \eqref{YM-action_T-V} takes the form
\begin{align}\label{slow-time-YM-action}
    S_{\rm YM}(\Acal)=\dfrac{1}{\hat{e}^2}\int_{\Ical_\ve}\diff\tau\;\left(\Tcal_\ve(\Acal)-\dfrac{1}{\ve^2}\Vcal(\Acal)\right)\ ,
\end{align}
where $\hat{e}^2=e^2/\ve$ is a rescaling of the gauge coupling. The potential energy $\Vcal(\Acal)$ is unaffected by this. However the kinetic energy becomes
\begin{align}\label{slow-time-kinetic-energy}
    \Tcal_\ve(\Acal)=-\dfrac{1}{2}\int_{S^3}\diff V_3\;\tr(\Fcal_{\tau a}\Fcal_{\tau a})\ ,
\end{align}
where $\Fcal_{\tau a}=\sfrac{1}{\ve}\Fcal_{ta}$ which can be recognized under the substitution $\Acal_t=\ve\Acal_\tau$ and $\pa_t=\ve\pa_\tau$ into \eqref{curvature-1}. Similarly, the term $S_j(\Acal)$ of the extended action rescales to
\begin{equation}
    S_j(\Acal)=\dfrac{1}{\hat{e}^2}\int_{\Ical_\ve\times S^3}\diff\tau\wedge \diff V_3\;\tr(\Acal_\tau j_\tau),
\end{equation}
where $j_\tau=\sfrac{1}{\ve} j_t$.
\smallskip

\noindent {\bf Energy.} The conserved energy density of Yang--Mills configurations (with or without a static source) is
\begin{equation}\label{energy-density}
\Ecal_t \= -\sfrac{1}{4e^2}\tr (2\Fcal_{ta}\,\Fcal_{ta} +  \Fcal_{ab}\,\Fcal_{ab})\ .
\end{equation}
One can also introduce the energy
\begin{equation}\label{energy}
E_t = \int_{S^3}\diff V_3\ {\cal E}_t=\frac{1}{e^2}\bigl(\Tcal(\Acal) + \Vcal(\Acal)\bigr)\ .
\end{equation}
Both $\Ecal_t$ and $E_t$ are positive-semidefinite and invariant under the group $\Gcal^0$ of gauge
transformations and the group of physical symmetries $\Gcal_{S^3}$.

The expressions \eqref{energy-density} and \eqref{energy} may instead be considered in slow time via the equivalence $\Ecal_t\,\diff t=\Ecal_\tau\,\diff\tau$ and $E_t\,\diff t=E_\tau\,\diff\tau$, where
\begin{equation}\label{slow-time-energy-density}
\Ecal_\tau \= -\frac{1}{4\hat e^2}\tr (2\Fcal_{\tau a}\,\Fcal_{\tau a} + \frac{1}{\ve^2} \Fcal_{ab}\,\Fcal_{ab})\ ,
\end{equation}
and
\begin{equation}\label{slow-time-energy}
E_\tau = \frac{1}{\hat e^2}\bigl(\Tcal_\ve(\Acal)+ \frac{1}{\ve^2}\,\Vcal(\Acal)\bigr)\ .
\end{equation}
From \eqref{slow-time-energy} we see that the case $0<\ve\ll 1$ corresponds to the low-energy limit if in this limit $\Vcal(A)\to 0$.

\smallskip

\noindent {\bf Yang--Mills equations on $\Ical_\ve\times S^3$}. The Euler-Lagrange equations for the rescaled version of the action $S=S_{\rm YM}+S_j$ are
\begin{equation}\label{YM-eqn}
\nabla_\tau\Fcal_{a\tau}=\frac{1}{\ve^2}\,\left(\nabla_b\Fcal_{ab}+\ve_{bc}^a\Fcal_{bc}\right)\ ,
\end{equation}
\begin{equation}\label{Gauss-law-slow}
\nabla_a\Fcal_{a\tau}=j_\tau\ ,
\end{equation}
with the covariant derivatives defined by \eqref{covariant-derivatives}, and $\nabla_\tau=\pa_\tau+[\Acal_\tau,\cdot]$.

\smallskip

\noindent {\bf Localization at $\Vcal(\Acal)=0$ for $\ve\to 0$.}
In the low-energy limit $\ve\to0$, the Yang--Mills equations \eqref{YM-eqn} are split into two sets of equations,
\begin{equation}\label{YM-eqn1}
\nabla_\tau\Fcal_{\tau a}=0
\end{equation}
and
\begin{equation}\label{YM-eqn2}
\nabla_b\Fcal_{ab}+\ve_{bc}^a\Fcal_{bc}=0\ .
\end{equation}
From the variation $\de S$ we observe the same splitting, but from the action \eqref{slow-time-YM-action} we see that
$\ve^{-2} \Vcal(\Acal)$ becomes highly peaked about $\Vcal(\Acal)=0$ as $\ve^2\to 0$. Hence, the partition function
of quantum Yang--Mills theory will be dominated by the zeros of $\Vcal$ in the low-energy limit $\ve^2\to0$, namely solutions to
\begin{equation}\label{vacuum}
\Vcal(\Acal)=0\ .
\end{equation}
This is also supported by the known fact
that flat connections on $S^3$ realize minima \eqref{vacuum} of the potential $\Vcal(\Acal)$, and they are the only {\it stable}
solutions of Yang--Mills theory on $S^3$ \cite{Stern}. Hence, we consider static connections in order to study the moduli space of
vacua and the dynamics of Yang--Mills theory on $\Ical_\ve \times S^3$ as slow motion on the space of static connections.

\bigskip

\section{Moduli space of static Yang--Mills connections and vacua on $\Ical\times S^3$ }\label{sec-Moduli-static}

\noindent {\bf Static connections and reduced gauge group.} Let $\Abb_{S^3}\subset\Abb_M$ denote the space of static connections, where by \textit{static connection}, we mean one such that $\Acal_t=0$ and $\pa_t\Acal_a=0$. The subgroup of the gauge group $\Gcal^0$ which preserves the static connections is trivial, but importantly, the symmetry group \eqref{GS3}, $\Gcal_{S^3}$, acts on $\Abb_{S^3}$ non-trivially in the obvious way via \eqref{gt}.

\smallskip

\noindent {\bf Flat connections on $S^3$.} For static connections $\Acal_t=0$ and $\pa_t\Acal_a=0$ the ``kinetic energy" \eqref{kinetic-energy} vanishes.
Then the total energy \eqref{slow-time-energy} vanishes if $\Vcal(\Acal)=0$. Since our bundle is trivial, and $\pi_1(S^3)=0$, this is achieved by
\begin{equation}\label{flat-S^3}
\Fcal_{ab} = 0\quad \Leftrightarrow\quad \Acal_a=g^{-1}L_a\,g\ ,
\end{equation}
where $g\in\Gcal_{S^3}$. In other words, the set of flat connections on $S^3$ is the orbit of the trivial connection $\Acal_a=0$ under the action \eqref{gt} of $\Gcal_{S^3}$. Note that this representation is not unique for all $g\in\Gcal_{S^3}$; $g,g'\in\Gcal_{S^3}$ define the same flat connection \eqref{flat-S^3} if and only if $g'=kg$ for some constant $k\in G$. Hence, the true moduli space\footnote{Here we use the term moduli space even though we are quotienting by the trivial group.} $\Mcal_{\rm vac}$ of flat connections on $S^3$ is the set of left cosets $\Gcal_{S^3}/G$. This principal homogeneous space may be identified as
\begin{equation}\label{Mvac}
\Mcal_{\rm vac}=\Gcal_{S^3}/G\cong\Gcal_{S^3}^n\ ,
\end{equation}
the group of based maps \cite{PS}, i.e. the normal subgroup $\Gcal_{S^3}^n$ of $\Gcal_{S^3}$ given by the kernel of the evaluation map
\begin{equation}\label{eval-map}
    \Gcal_{S^3}\ \to \ G,\quad g\mapsto g(n)\ ,
\end{equation}
where $n$ is a point on $S^3$, e.g. the north pole.

\smallskip

\noindent {\bf Tangent space.} As discussed in Section \ref{sec-GT-on-M}, the tangent spaces to $T_\Acal\Abb_{S^3}$ are identified with the spaces of one-forms $\Lambda^1(S^3$, Ad$P$). Restricting to the subspace $\Mcal_{\rm vac}\subset\Abb_{S^3}$, tangent vectors are additionally required to solve the linearized flatness equations $\nabla_\Acal\;\delta\Acal=0$, which in components with respect to the frame $e^a$ of $T^\ast S^3$ reads
\begin{equation}\label{Lin-flat}
\nabla_a\,\de\Acal_b - \nabla_b\,\de\Acal_a - 2\ve^c_{ab}\de\Acal_c =0\ .
\end{equation}
Since every element of $\Mcal_{\rm vac}$ corresponds to a point in the orbit of the action of $\Gcal_{S^3}^n$ on $\Acal_a=0$, every tangent vector $\delta\Acal$ will likewise correspond to tangent vectors induced by the infinitesimal action, i.e. of the form $\delta\Acal=\nabla_\Acal\;\Psi=\nabla_a\Psi\;e^a$, for $\Psi\in\mathrm{Lie}(\Gcal_{S^3}^n)$. It is clear that vectors of this form solve \eqref{Lin-flat}. It is important to stress again that we do not consider these as gauge variations to be removed; since our gauge group is trivial, these are precisely the tangent vectors that we are interested in.

\smallskip

\noindent {\bf Coordinate frame for $T\Mcal_{\rm vac}$.} Let $X^\alpha$, $\alpha=1,2,\ldots$, denote a set of local coordinates on $\Mcal_{\rm vac}$, with partial derivatives $\pa_\alpha\equiv\sfrac{\pa}{\pa X^\alpha}$. As discussed above, every flat connection $\Acal\in\Mcal_{\rm vac}$ is determined uniquely by a choice of based map $g\in\Gcal_{S^3}^n$. The function $g\in\Gcal_{S^3}^n$, and hence $\Acal_a$ depends on the coordinates $X^\alpha$. It is clear that $g^{-1}\pa_\alpha g\in\mathrm{Lie}(\Gcal_{S^3}^n)$, and one finds
\begin{equation}
    \pa_\alpha\Acal_a=\nabla_a(g^{-1}\pa_\alpha g)\ ,\label{tangent-vec}
\end{equation}
i.e. a general tangent vector to $\Mcal_{\rm vac}$. Therefore, given a flat connection $\Acal_a=g^{-1}L_ag$ with $g\in\Gcal_{S^3}^n$, the objects $\de_\alpha\Acal_a:=\pa_\alpha\Acal_a$ define a coordinate basis for $T_\Acal\Mcal_{\rm vac}$.

\smallskip

\noindent {\bf Metric.} Restricting the inner product \eqref{metric-1forms-S^3} on $\Abb_{S^3}$ to the subspace
$\Mcal_{\rm vac}\subset\Abb_{S^3}$ provides a metric $\Gbb =(G_{\al\be})$ in terms of the coordinates $X^\alpha$ for the moduli space $\Mcal_{\rm vac}\cong\Gcal^n_{S^3}$
of static vacua, namely
\begin{equation}\label{metric-Mvac}
G_{\al\be}=-\int_{S^3}\, \diff V_3\, \tr(\de_\al\, \Acal_a\ \de_\be\, \Acal_a)\ .
\end{equation}
We also consider the Levi-Civita connection on $T\Mcal_{\rm vac}$, with Christoffel symbols written in terms of \eqref{metric-Mvac} as
\begin{equation}\label{Christoffel-symbols1}
\Gamma^\ga_{\al\be}=\sfrac12\, G^{\ga\vk}(\pa_\al\, G_{\be\vk}+ \pa_\be\, G_{\al\vk}-\pa_\vk\, G_{\al\be} )\ .
\end{equation}
It is straightforward to show that these may be written in the form
\begin{equation}\label{Christoffel-symbols-int-version}
\Gamma^\ga_{\al\be}=G^{\ga\vk}\int_{S^3}\, \diff V_3\,
\tr(\de_\vk\, \Acal_a\, \pa_\al\, \de_\be\, \Acal_a)\ .
\end{equation}
One can also introduce the Riemann curvature tensor, Ricci tensor etc. but these are not relevant for our purposes.

\smallskip

\noindent {\bf Full space of static Yang--Mills.} Flat connections \eqref{flat-S^3} on $S^3$ realize the absolute minima
$\Vcal(\Acal)=0$ of the potential \eqref{potential-energy} which is exactly the Euclidean action of Yang--Mills theory on $S^3$. The equations of motion for Yang--Mills theory on $S^3$ are \eqref{YM-eqn2}. Flat connections are the
{\it only stable} solutions, and there are no topologically nontrivial solutions
\cite{Stern}. There is at least one unstable solution having the form
\begin{equation}\label{unstable-YM-soln-S^3}
\Acal^0 = \sfrac12\, g_1^{-1}\, \diff g_1 = \sfrac12\, g_1^{-1}L_a g_1\,e^a=\sfrac12\, e^aI_a\ ,
\end{equation}
where $g_1: S^3\to G$ is a map of degree one, and $\{I_a\}$ are generators of the group SU(2)$\,\subseteq G$
forming a part of the generators of $G$.

For $G=\,$SU(2), the solution \eqref{unstable-YM-soln-S^3} is the standard metric connection without torsion (Levi-Civita connection) on the frame bundle over $S^3$. Similarly, for $G=\,$SU(2), the flat connections
\begin{equation}\label{pm-flat-connections}
\Acal^- = 0\und \Acal^+ =g_1^{-1}L_a g_1\,e^a=e^aI_a
\end{equation}
are the metric compatible connections with ($\mp$) torsion trivializing the SU(2)-frame bundle over $S^3$. Obviously,
$\Acal^+$ is simply the transformation of $\Acal^-$ via \eqref{gt} with $g_1\in\Gcal^n_{S^3}$, and thus belong to the orbit
$\Gcal^n_{S^3}\subset \Abb_{S^3}$ passing through the point $0=\Acal^-\in \Abb_{S^3}$. Similarly, the unstable solution
\eqref{unstable-YM-soln-S^3} can be transformed by the group $\Gcal^n_{S^3}$ into an infinite family of gauge-inequivalent
solutions from the viewpoint of gauge theory on de Sitter space dS$_4$.

Recall that for Yang--Mills theory on $\Ical\times S^3$ with trivial holonomy (which is our present case of interest), we may consider allowed variations of $\Acal_a$ on $S^3$ of the form \eqref{allowed-variations} without the emergence of non-zero boundary terms. In this case the only admissible connections on $S^3$ are the SU(2)-equivariant family
\begin{equation}\label{admissable-connection-S^3}
\Acal_a=\vk\, e^aI_a\ ,
\end{equation}
where $\vk\in\R$ is a free parameter and $\{e^a\}$ are the left-invariant one-forms \eqref{l-i-1forms} on $S^3$. It is well known
(see e.g. \cite{ILP2,Kub}) that the family \eqref{admissable-connection-S^3} allows only three solutions of the Yang--Mills equations
on $S^3$: specifically with $\vk=\sfrac12$, $\vk =0$, or $\vk=1$. These three solutions are given in \eqref{unstable-YM-soln-S^3} and \eqref{pm-flat-connections}.
Thus, for the considered boundary conditions, the moduli space of Yang--Mills connections on $S^3$ is a disjoint union
of two group manifolds $\Gcal^n_{S^3}$,
\begin{equation}\label{Full-YM-vacuum}
\Mcal_{\rm full}^{S^3} =\Gcal^n_{S^3}\cup \Gcal^n_{S^3}\,,
\end{equation}
determined by the action \eqref{gt} of $\Gcal^n_{S^3}$ on the solutions \eqref{unstable-YM-soln-S^3}-\eqref{pm-flat-connections}. However, solutions generated from \eqref{unstable-YM-soln-S^3} are saddle points of the potential $\Vcal(\Acal)$ and, due to
instability, for any small velocity $\pa_t\Acal_a$ they will oscillate around the minima, i.e. solutions to $\Vcal(\Acal)=0$. That is why in the
low-energy limit  we will consider only the moduli space \eqref{Mvac}. This argument complements the previous one from the end of Section \ref{sec-YM-on-M}, based on localization at $\Vcal(\Acal)=0$.

\bigskip

\section{An adiabatic limit of Yang--Mills theory on $\Ical\times S^3$}\label{sec-Low-energy}

\smallskip

\noindent {\bf Slow motion on $\Abb_{S^3}$.} Having established the description of the static configuration space
$\Abb_{S^3}$ and the moduli space $\Mcal_{\rm vac}$ of static vacua of Yang--Mills theory on $\Ical\times S^3$,
we return to dynamic gauge fields depending on $t\in \Ical$. Recall that we may think of these as a paths
\begin{equation}
    \Acal:\Ical\to\Abb_{S^3},\quad t\mapsto\Acal(t)\ ,
\end{equation}
where the component $\Acal_t$ of $\Acal$ may be determined by resolving the relaxed Gauss law $\nabla_a\Fcal_{at}=j_t$, which encodes the additional physical degrees of freedom imposed by the framing of the bundle $P$.

Consider now a family of paths $\Acal (\ve , t)\in\Abb_{S^3}$ depending on a small parameter $\ve >0$ and
such that the kinetic energy \eqref{kinetic-energy} is $\Tcal(\Acal (\ve ,t))\approx\ve^2\,\ll\,1$. For small $\ve$ the
dynamic solutions $\Acal (\ve , t, x)$, $x\in S^3$, are close to the static solutions from
$\Mcal_{\rm vac}\subset\Abb_{S^3}$ and in the limit $\ve\to 0$ they converge to a point $\Acal\in\Mcal_{\rm vac}$.
However, an established adiabatic method proposes a more refined approach, which yields a
geodesic on $\Mcal_{\rm vac}$ instead of a point. Furthermore, one expects this geodesic to be close to the path in $\Abb_{S^3}$
defined by a true dynamic solution $\Acal (\ve ,t)$ for small $\ve$, as has been confirmed in a variety of related cases \cite{Man, Serg, Stu, Uhl}.

\smallskip

\noindent {\bf Adiabatic approach.} In the case of Yang--Mills theory on $\Ical\times S^3$, the adiabatic
approach to describing dynamic solutions implies the following steps:
\begin{enumerate}
    \item One considers static solutions, i.e. solutions of the Yang--Mills equations \eqref{YM-eqn2} on $S^3$, and describes the moduli space $\Mcal_{\rm vac}$ of their vacuum configurations.
    \item One introduces ``slow time" $\tau:={\ve} t$, rescaling the interval to $\Ical_\ve$, rewrites the action and energy functionals by using $\tau$ and shows that the limit $\ve\to 0$ corresponds to the low-energy limit of Yang--Mills theory on $\Ical_\ve \times S^3$.
    \item One allows the collective coordinates $X^\alpha$ on the moduli space $\Mcal_{\rm vac}$ to depend on $\tau$, and assumes that the connection $\Acal$ depends on $\tau$ only via the coordinates $X^\alpha$, i.e. $\Acal = \Acal (X^\al (\tau ),x)$. One then substitutes $\Acal (X^\al (\tau ),x)$ into the rescaled Yang--Mills action \eqref{slow-time-YM-action}.
    \item One performs the small-$\ve$ limit in the action and in the corresponding Yang--Mills equations. Then one shows that Yang--Mills theory on $\Ical_\ve\times S^3$ reduces to a sigma model describing maps from $\Ical_\ve$ into the moduli space $\Mcal_{\rm vac}$ of vacua. The Yang--Mills equations in this case reduce to the equations for geodesics on the manifold $\Mcal_{\rm vac}$. 
\end{enumerate}

\smallskip

\noindent {\bf Moduli-space approximation.} In previous sections we have executed steps 1. and 2. of the
adiabatic approach outlined above, and we described the moduli space $\Mcal_{\rm vac}\cong\Gcal^n_{S^3}$ of static vacua of
Yang--Mills theory on $\Ical_\ve\times S^3$. Now we return to full Yang--Mills theory. According to step 3.,
we let the moduli parameters $X=\{X^\al\}$ of $\Mcal_{\rm vac}$
define a map
\begin{equation}\label{curve-in-Mvac}
X\ :\quad\Ical_\ve\to\Mcal_{\rm vac}
\end{equation}
from $\Ical_\ve$ to $\Mcal_{\rm vac}=\Gcal^n_{S^3}$. Thus,
$X^\al(\tau)$ may be considered as dynamical fields which capture the $\tau$-dependence
of ``slow" full Yang--Mills solutions. The low-energy effective action for $X^\al$ is derived by the leading term of the Yang--Mills action \eqref{slow-time-YM-action} in the expansion
\begin{equation}\label{low-energy-exp}
\Acal = \Acal (X^\al(\tau), x) + O(\ve)\ ,
\end{equation}
where the first term depends on $\tau\in\Ical_\ve$ only via the coordinates $X^\al\in\Mcal_{\rm vac}$ \cite{ Har, Stu, HS, Gau}.
For small $\ve\ll 1$, all terms in \eqref{low-energy-exp} beyond the first one are discarded. By substituting the leading term of \eqref{low-energy-exp} into the
action \eqref{slow-time-YM-action}, one obtains an effective field theory describing small fluctuations around the vacuum manifold
$\Mcal_{\rm vac}$. Note that $\Mcal_{\rm vac}$ contains all topological sectors, i.e. the connected components of $\Gcal_{S^3}^n$ corresponding to the distinct homotopy classes, and therefore the
consideration is not a perturbative one.

\smallskip

\noindent \textbf{Effective action.} For $\Acal\in \Mcal_{\rm vac}$, $\Vcal(\Acal)=0$, hence the effective action is determined solely by the kinetic term. Since we consider trivial holonomy, we may choose the gauge $\Acal_\tau=0$, so that, in the adiabatic limit $\ve\to0$, we have the electric field
\begin{equation}
    \Fcal_{\tau a}=\pa_\tau\Acal_a=(\pa_\tau X^\alpha)\pa_\alpha\Acal_a\ .\label{curvature-tau-trivial-holonomy}
\end{equation}
Upon substitution into \eqref{slow-time-YM-action}, we obtain
\begin{equation}\label{effective-action}
S=\frac{1}{\hat e^2}\, \int_{\Ical_\ve} \diff\tau\, G_{\al\be}\pa_\tau X^\al\pa_\tau X^\be\ ,\quad\Rightarrow\quad S_{\rm eff}=\frac{1}{e^2}\, \int_{\Ical} \diff t\, G_{\al\be}\pa_t X^\al\pa_t X^\be\ ,
\end{equation}
where we have used the formula \eqref{metric-Mvac} for the metric components $G_{\al\be}$ on $\Mcal_{\rm vac}$ with respect to the coordinate basis $\de_\alpha\Acal_a=\pa_\alpha\Acal_a$ of $T_\Acal\Mcal_{\rm vac}$. Thus, in the case of trivial holonomy, the Yang--Mills action \eqref{slow-time-YM-action}
reduces for $\ve\ll 1$ to the action of a nonlinear sigma model on ${\Ical_\ve}$, with target space $\Mcal_{\rm vac}=
C^\infty(S^3, G)/G=\Gcal^n_{S^3}$, which is an infinite-dimensional group manifold.

\smallskip

\noindent {\bf Geodesics on $\Mcal_{\rm vac}$.} Recall that for small $\ve\ll 1$ the Yang--Mills equations
on ${\Ical_\ve}\times S^3$ are reduced to the equations \eqref{Gauss-law-slow} and \eqref{YM-eqn1}-\eqref{YM-eqn2}. Also recall that in the case of trivial holonomy, the role of the field $\lambda$ may be viewed as a Lagrange multiplier for the gauge choice $\Acal_\tau=0$, and that the relaxed Gauss law \eqref{relaxed-Gauss-law} determines the Lagrange multiplier $\lambda$, i.e. $\lambda=\nabla_a\Fcal_{a\tau}$, but is not a dynamical equation of motion. Since $\Fcal_{ab}=0$ for $\Acal\in\Mcal_{\rm vac}$, so \eqref{YM-eqn2} is satisfied, there remains only \eqref{YM-eqn1}, namely\footnote{Here we make the replacement $\tau\mapsto t$ so that everything is defined on a fixed-length interval.}
\begin{equation}\label{YM-eqn2-revisit}
\nabla_t\Fcal_{ta}=0\ ,\quad a=1,2,3\ .
\end{equation}
Substituting \eqref{curvature-tau-trivial-holonomy} into \eqref{YM-eqn2-revisit} with the gauge choice $\Acal_t=0$, we obtain
\begin{equation}\label{EOM1}
(\pa^2_t X^\al)\,\de_\al  \Acal_a +\pa_t X^\be\pa_t X^\al\pa_\al\de_\be \Acal_a =0\ .
\end{equation}
This implies
\begin{equation}\label{EOM2}
\int_{S^3}\diff V_3\,\tr\left((\pa^2_t X^\al)\,\de_\vk\Acal_a\de_\al  \Acal_a +\pa_t X^\be\pa_t X^\al\de_\vk\Acal_a\pa_\al\de_\be \Acal_a\right)=0\ ,
\end{equation}
which, as we see from \eqref{metric-Mvac} and \eqref{Christoffel-symbols-int-version}, leads to
\begin{equation}\label{geodesic-eqn}
\pa^2_t X^\al + \Gamma^\al_{\be\ga}\,\pa_t X^\be\pa_t X^\gamma =0\ .
\end{equation}
Equations \eqref{geodesic-eqn} for $\alpha=1,2,\ldots$ are the equations of a geodesic $X=(X^\alpha):\Ical\to\Mcal_{\rm vac}$ on the moduli space
$\Mcal_{\rm vac}$ of vacua. They are the Euler-Lagrange equations for the action $S_{\rm eff}$ in \eqref{effective-action}.

\smallskip

\noindent {\bf Adiabatic limit of unstable solutions.} Note that one can also consider small fluctuations around the infinite-dimensional
manifold $\Gcal^n_{S^3}$ of static solutions to the full Yang--Mills equations on $S^3$ generated by the unstable solution \eqref{unstable-YM-soln-S^3}. Since the moduli space of these solutions is the same as the moduli space
$\Mcal_{\rm vac}$ of vacua, we will arrive to the same effective field theory \eqref{effective-action} which describes
small fluctuations around solutions of the form
\begin{equation}\label{general-unstable-soln}
\Acal^0=\sfrac12\,(g_1 g)^{-1}\diff\, (g_1 g) + \sfrac12\,g^{-1}\diff g\ ,\text{ for }\ g\in\Gcal^n_{S^3}\ .
\end{equation}
However, the moduli space \eqref{Full-YM-vacuum} is a disjoint union only for static solutions.
For the solutions \eqref{general-unstable-soln} the potential energy $\Vcal(\Acal^0)=\sfrac38$ (see e.g. \cite{ILP2}), and they are unstable.
Hence, after switching on a dependence on time this solution will oscillate around the vacuum solutions for any small kinetic energy $\Tcal(\Acal^0(\tau))$.

\smallskip

\noindent {\bf The Gauss law revisited.} Throughout this paper we have emphasized that one should not impose the Gauss law constraint \eqref{Gauss-law} when considering Yang--Mills theory on framed bundles over $\Ical\times S^3$, as it kills dynamical degrees of freedom. This property is also manifest when considering the adiabatic limit. Indeed, regardless of whether we consider trivial, or non-trivial holonomy, we could determine the non-dynamical variable $\Acal_t$ using the Gauss law as in the argument \eqref{resolving-Gauss-law}, but set $\lambda=j_t=0$. In the case of flat connections, we know that $\pa_t\Acal_a=\nabla_a(g^{-1}\pa_tg)$, and so \eqref{resolving-Gauss-law} reduces further to
\begin{equation}
    \nabla^2(\Acal_t-g^{-1}\pa_tg)=0\ ,
\end{equation}
i.e. $\Acal_t=g^{-1}\pa_tg+\varphi$, where $\varphi\in\ker\nabla^2\subset C^\infty(S^3,\mathfrak{g})$ is (covariantly) harmonic. However, since $S^3$ is a closed manifold, it is well-known that the only (covariantly) harmonic functions are (covariantly) constant, i.e. $\nabla_a\varphi=0$. Plugging this back into the formula \eqref{curvature-1} for the electric field yields $\Fcal_{ta}=-\nabla_a\varphi=0$. Therefore, this adiabatic limit is trivial when the Gauss law \eqref{Gauss-law} is imposed.

More generally, in the case of irreducible connections, in which case $\nabla^2$ is invertible, one can show that the resolution \eqref{resolving-Gauss-law} of the Gauss law constraint \eqref{Gauss-law} leads to
\begin{equation}
    \Fcal_{ta}=(\Pi\pa_t\Acal)_a=\left(\delta_{ab}-\nabla_a\nabla^{-2}\nabla_b\right)\pa_t\Acal_b\ ,
\end{equation}
so that the electric field is the image under the covariant transversal projection $\Pi:\Abb_{S^3}\to\Abb_{S^3}/\Gcal_{S^3}$ of $\pa_t\Acal$. In particular, this means that $\Fcal_{ta}$ is formed only from dynamical variables $\Acal_a$ in the moduli space $\Abb_{S^3}/\Gcal_{S^3}$, i.e. the Gauss law has the effect of quotienting out the $\Gcal_{S^3}$ degrees of freedom, as explained in the discussions in both Section \ref{sec-GT-on-M} and Section \ref{sec-YM-on-M}. Since the space of flat connections in $\Abb_{S^3}/\Gcal_{S^3}$ consists of a point, namely $\Acal_a=0$, from this perspective the above result is clear.

\smallskip

\noindent {\bf Integrability.} Since $\Mcal_{\rm vac}$ is a Lie group $\Gcal^n_{S^3}$, one can construct geodesics as
one-parameter subgroups of $\Gcal^n_{S^3}$. If the metric \eqref{metric-Mvac} on the group $\Gcal^n_{S^3}$ is bi-invariant,
which is pretty likely, then all geodesics are one-parameter subgroups.\footnote{ For a brief review see e.g.
\cite{Ale} and references therein.} Hence, the low-energy limit of Yang--Mills theory on de Sitter space
dS$_4$, with trivial holonomy, is the integrable principal chiral model in one dimension with the group $\Gcal^n_{S^3}$ as a target space.
From the implicit function theorem it follows that for any approximate solution $\Acal(\ve =0)$ defined by a geodesic
\eqref{geodesic-eqn} on $\Mcal_{\rm vac}$, there exist nearby solutions $\Acal (\ve>0)$ of the Yang--Mills equations on
dS$_4$ for $\ve$ sufficiently small. It is therefore reasonable to conjecture that the moduli space of all geodesics on
$\Mcal_{\rm vac}$ is bijective to the moduli space $\Mcal_{\rm YM}(\mathrm{Id},\lambda)$ of trivial holonomy solutions to the Yang--Mills equations on dS$_4$. Regardless,
it is worth exploring further relations of Yang--Mills theory on dS$_4$ and one-dimensional principal chiral models
from the viewpoint of integrability, in the low-energy limit, and beyond.

\bigskip

\section{Conclusions}\label{sec-concl}

\noindent
By exploiting the conformal invariance of Yang--Mills theory in four dimensions, we reduced Yang--Mills theory on de Sitter space dS$_4$ in a certain adiabatic limit to a
one-dimensional principal chiral model with the moduli space $\Mcal_{\rm vac}$ of static gauge vacua as a
target space, where in particular we identified $\Mcal_{\rm vac}$ with the infinite-dimensional Lie group $\Gcal^n_{S^3}\cong C^\infty (S^3, G)/G$.
This principal chiral model captures the low energy dynamics of Yang--Mills theory on dS$_4$. This example is a demonstration of a more general idea:
in the presence of a boundary, the group of gauge transformations becomes smaller, which yields additional
degrees of freedom localized at the boundary. In our case, we described an infinite-dimensional dynamical symmetry group
acting on the boundary states of Yang--Mills theory on dS$_4$. This group is responsible for the appearance of
an infinite-dimensional manifold of inequivalent ground states - the classical moduli space $\Mcal_{\rm vac}$ of vacua.
Consideration of such dynamical symmetry groups is important in quantum  Yang--Mills theory on Minkowski space $\R^{3,1}$, and it would be interesting to study further the role of these dynamical symmetry groups in the context of Yang--Mills theory
on de Sitter space dS$_4$. We also remind the reader that we only considered connections with trivial holonomy. Much of the formalism presented in this paper already presents a significant departure from the standard analysis for Yang--Mills theory on Minkowski space $\R^{3,1}$, and the addition of non-trivial holonomy introduces further subtle complexities which are reserved for consideration in a further work.

An important consequence of our considerations has been to highlight the role of the Gauss law constraint \eqref{Gauss-law} 
and, in particular, when it can and cannot be imposed. This is directly related to the choices of boundary conditions (see e.g. \cite{Bal}). 
In our case, the framing over the boundary introduces non-dynamical degrees of freedom, and na\"ive imposition of constraints such as the Gauss law \eqref{Gauss-law} leads to the loss of these data. 
Our resolution to this problem was to pair the constraints with these non-dynamical degrees of freedom, motivated both by a variational (Lagrangian) and symplectic geometry (Hamiltonian) approach to Yang--Mills theory. Our work motivates the need for a deeper understanding of the relationships in Yang--Mills theory between framing of bundles, allowed variations, constraints, and beyond, and some directions we are currently exploring, alongside the non-trivial holonomy case mentioned above, are questions relating to framing over more general submanifolds of temporal boundaries and how this relates both to the formalism of the relaxed Gauss law \eqref{relaxed-Gauss-law}, and adiabatic limits presented in this paper. Finally, although our focus has been on de Sitter space (and ultimately the cylinder $\Ical\times S^3$), the formulation we have discussed here is applicable to a variety of other examples by framing over a temporal boundary. For example, a simple generalization of our results would apply to any spacetime with conformal structure of the form $I\times\Sigma$ with $I\subset\R$ a timelike interval, and $\Sigma$ a spacelike closed manifold, by introducing a framing over $\pa I\times\Sigma$.

\bigskip
\noindent {\bf Acknowledgements}

\noindent
This work was partially supported by the Deutsche Forschungsgemeinschaft grant LE~838/19. 
E\c SK gratefully acknowledges the support of the Riemann Fellowship at Leibniz Universit\"at Hannover, and also the Research Fund of the Middle
East Technical University, Project Number: DOSAP-B-105-2021-10763.


\newpage

\begin{thebibliography}{99}
\bibitem{Strom}
  A.~Strominger,
  {\it Lectures on the infrared structure of gravity and gauge theory,}\\
  Princeton University Press, Princeton, 2018.

\bibitem{Kap}
  D.~Kapec, M.~Perry, A.M.~Raclariu and A.~Strominger,
  ``Infrared divergences in QED, revisited,''
  Phys.\ Rev.\ D {\bf 96} (2017) 085002
  [arXiv:1705.04311 [hep-th]].
  
  \bibitem{Ser}
  A.~Seraj and D.~Van den Bleeken,
  ``Strolling along gauge theory vacua,''\\
  JHEP {\bf 08} (2017) 127
  [arXiv:1707.00006 [hep-th]].
  
  \bibitem{Hen}
  M.~Henneaux and C.~Troessaert,
  ``Asymptotic symmetries of electromagnetism at spatial infinity,''
  JHEP {\bf 05} (2018) 137
  [arXiv:1803.10194 [hep-th]].
  
  \bibitem{Hos}
  V.~Hosseinzadeh, A.~Seraj and M.M.~Sheikh-Jabbari,
  ``Soft charges and electric-magnetic duality,''
  JHEP {\bf 08} (2018) 102
  [arXiv:1806.01901 [hep-th]].
  
  \bibitem{Stie}
  S.~Stieberger and T.R.~Taylor,
  ``Symmetries of celestial amplitudes,''\\
  Phys.\ Lett.\ B {\bf 793} (2019) 141
  [arXiv:1812.01080 [hep-th]].
  
 \bibitem{Him}
  E.~Himwich and A.~Strominger,
  ``Celestial current algebra from low’s subleading soft theorem,''
  Phys.\ Rev.\ D {\bf 100} (2019)   065001
  [arXiv:1901.01622 [hep-th]].
 
 \bibitem{DiP}
  L.~Di Pietro, D.~Gaiotto, E.~Lauria and J.~Wu,
  ``3$d$ Abelian gauge theories at the boundary,''
  JHEP {\bf 05} (2019) 091
  [arXiv:1902.09567 [hep-th]].
 
 \bibitem{Pate}
  M.~Pate, A.M.~Raclariu and A.~Strominger,
  ``Conformally soft theorem in gauge theory,''\\
  Phys.\ Rev.\ D {\bf 100}, (2019) 085017
  [arXiv:1904.10831 [hep-th]].
 
 \bibitem{Bal}
  A.P.~Balachandran, V.P.~Nair and S.~Vaidya,
  ``Aspects of boundary conditions for non-Abelian gauge theories,''
  Phys.\ Rev.\ D {\bf 100} (2019)  045001
  [arXiv:1905.00926 [hep-th]].
 
 \bibitem{Gon}
  R.~Gonzo, T.~Mc~Loughlin, D.~Medrano and A.~Spiering,
  ``Asymptotic charges and coherent states in QCD,''
  arXiv:1906.11763 [hep-th].
  
  \bibitem{Gid}
  S.B.~Giddings,
  ``Generalized asymptotics for gauge fields,''\\
  JHEP {\bf 2019}, 66 (2019)
  [arXiv:1907.06644 [hep-th]].
 
 \bibitem{Str}
  A.~Strominger,
  ``Asymptotic symmetries of Yang--Mills theory,''\\
  JHEP {\bf 07} (2014) 151
  [arXiv:1308.0589 [hep-th]].
  
  \bibitem{HennTroes}
  M.~Henneaux, and C.~Troessaert, ``A note on electric-magnetic duality and soft charges,''\\
  JHEP {\bf 06} (2020) 081
  [arXiv:2004.05668 [hep-th]].
  
  \bibitem{DonPasPuh}
  L.~Donnay, S.~Pasterski, and A.~Puhm,
  ``Asymptotic Symmetries and Celestial CFT,''\\
  JHEP {\bf 2020} 176 (2020) 
  [arxiv:2005.08990 [hep-th]].
  
  \bibitem{TanziGiulini}
  R.~Tanzi, and D.~Giulini, ``Asymptotic symmetries of Yang--Mills fields in Hamiltonian formulation,"
  JHEP {\bf 10} (2020) 094
  [arXiv:2006.07268 [hep-th]].
  
  \bibitem{He}
  T.~He, P.~Mitra, A.P.~Porfyriadis and A.~Strominger,
  ``New symmetries of massless QED,''
  JHEP {\bf 10} (2014) 112
  [arXiv:1407.3789 [hep-th]].
  
  \bibitem{HMS}
  T.~He, P.~Mitra and A.~Strominger,
  ``2D Kac-Moody symmetry of 4D Yang--Mills theory,''
  JHEP {\bf 10} (2016) 137
  [arXiv:1503.02663 [hep-th]].
  
  \bibitem{Donn}
  W.~Donnelly and L.~Freidel,
  ``Local subsystems in gauge theory and gravity,''\\
  JHEP {\bf 1609} (2016) 102
  [arXiv:1601.04744 [hep-th]].
  
   \bibitem{Gom}
  H.~Gomes, F.~Hopfm\"uller and A.~Riello,
  ``A unified geometric framework for boundary charges and dressings: non-Abelian theory and matter,''
  Nucl.\ Phys.\ B {\bf 941} (2019) 249
  [arXiv:1808.02074 [hep-th]].
  
  \bibitem{Mat}
  P.~Mathieu, A.~Schenkel, N.J.~Teh and L.~Wells,
  ``Homological perspective on edge modes in linear Yang--Mills theory,'' Lett.\ Math.\ Phys. (2020) 110
  [arXiv:1907.10651 [hep-th]].
  
  \bibitem{Riello}
  A.~Riello,
  ``Symplectic reduction of Yang--Mills theory with boundaries: from superselection sectors to edge modes, and back,''\\
  SciPost Phys. {\bf 10}, 125 (2021) [arXiv:2010.15894 [hep-th]].
  
 \bibitem{Pop}
  A.D.~Popov,
  ``Loop groups in Yang--Mills theory,''\\
  Phys.\ Lett.\ B {\bf 748} (2015) 439
  [arXiv:1505.06634 [hep-th]].
  
  \bibitem{ILP1}
  T.A.~Ivanova, O.~Lechtenfeld and A.D.~Popov,
  ``Solutions to Yang--Mills equations on four-dimensional de Sitter space,''
  Phys.\ Rev.\ Lett.\  {\bf 119} (2017)   061601
  [arXiv:1704.07456 [hep-th]].
  
  \bibitem{ILP2}
  T.A.~Ivanova, O.~Lechtenfeld and A.D.~Popov,
  ``Finite-action solutions of Yang--Mills equations on de Sitter dS$_{4}$ and anti-de Sitter AdS$_{4}$ spaces,''\\
  JHEP {\bf 11} (2017) 017
  [arXiv:1708.06361 [hep-th]].
  
 \bibitem{LZ}
  O.~Lechtenfeld and G.~Zhilin,
  ``A new construction of rational electromagnetic knots,''\\
  Phys.\ Lett.\ A {\bf 382} (2018) 1528
  [arXiv:1711.11144 [hep-th]].
  
 \bibitem{BV} 
  O.~Babelon and C.M.~Viallet,
  ``On the Riemannian geometry of the configuration space of gauge theories,''
  Commun.\ Math.\ Phys.\  {\bf 81} (1981) 515.

 \bibitem{Man}
  N.S.~Manton,
  ``A remark on the scattering of BPS monopoles,''
  Phys.\ Lett.\  {\bf 110B} (1982) 54.
  
 \bibitem{Har}
  J.A.~Harvey, G.W.~Moore and A.~Strominger,
  ``Reducing S-duality to T-duality,''\\
  Phys.\ Rev.\ D {\bf 52} (1995) 7161
  [hep-th/9501022].
 
 \bibitem{SW}
  N.~Seiberg and E.~Witten,
  ``Gauge dynamics and compactification to three-dimensions,''\\
  in *Saclay 1996, The mathematical beauty of physics* 333-366
  [hep-th/9607163].
  
 \bibitem{Serg}
 A.G.~Sergeev, ``Adiabatic limit in the Ginzburg--Landau and Seiberg--Witten equations,''\\
 Proc. Steklov Inst.\ Math. {\bf 289} (2015) 227.
 
 \bibitem{LP1}
  O.~Lechtenfeld and A.D.~Popov,
  ``Yang--Mills moduli space in the adiabatic limit,''\\
  J.\ Phys.\ A {\bf 48} (2015)  425401
  [arXiv:1505.05448 [hep-th]].
  
 \bibitem{Sei}
  N.~Seiberg,
  ``Exact results on the space of vacua of four-dimensional SUSY gauge theories,''
  Phys.\ Rev.\ D {\bf 49} (1994) 6857
  [hep-th/9402044].
  
  \bibitem{Bea}
  C.~Beasley and E.~Witten,
  ``New instanton effects in supersymmetric QCD,''\\
  JHEP {\bf 01} (2005) 056
  [hep-th/0409149].
  
  \bibitem{Eto}
  M.~Eto, Y.~Isozumi, M.~Nitta, K.~Ohashi and N.~Sakai,
  ``Solitons in supersymmetric gauge theories: moduli matrix approach,'' J.\ Phys. A {\bf 39} (2006) R315
 [hep-th/0607225].
  
  \bibitem{LP2}
  O.~Lechtenfeld and A.D.~Popov,
  ``Skyrme and Faddeev models in the low-energy limit of 4$d$ Yang--Mills--Higgs theories,''
  Nucl.\ Phys.\ B {\bf 945} (2019) 114675
  [arXiv:1808.08972 [hep-th]].
 
\bibitem{AH}
   M.F.~Atiyah and N.J.~Hitchin, {\it The geometry and dynamics of magnetic monopoles},\\ Princeton University Press, Princeton, 1988.
  
 \bibitem{MS}
N.~Manton and P.~Sutcliffe,
{\it Topological solitons},\\
Cambridge University Press, Cambridge, 2004. 
  
\bibitem{Don}
S.K.~Donaldson,
``Boundary value problems for Yang--Mills fields,''\\
J.\ Geom.\ Phys.\  {\bf 8} (1992) 89.

\bibitem{HE}
S.W.~Hawking and G.F.R.~Ellis,
{\it The large scale structure of space-time,}\\
Cambridge University Press, Cambridge, 1975.

\bibitem{Stern}
M.~Stern, ``Geometry of minimal energy of Yang--Mills connections,''\\
J. Differential Geometry  {\bf 86} (2010) 163.

\bibitem{HeHo}
J.E.~Hetrick and Y.~Hosotani, ``QED on a circle,''\\
Phys. Rev. D {\bf 38} (1988) 2621.

\bibitem{FadSlav}
    L.D.~Faddeev and Ao.A.~Slavnov, {\it Gauge Fields: Introduction to Quantum Theory,}\\
    Benjamin/Cummings, 1980

\bibitem{Kub}
  Y.A.~Kubyshin, V.O.~Malyshenko and D.~Marin Ricoy,
  ``Invariant connections with torsion on group manifolds and their application in Kaluza-Klein theories,''\\
  J.\ Math.\ Phys.\  {\bf 35} (1994) 310
  [gr-qc/9304047].
 
  \bibitem{Stu}
  D.~Stuart,
  ``The geodesic approximation for the Yang--Mills--Higgs equations,''\\
  Commun.\ Math.\ Phys.\  {\bf 166} (1994) 149.
  
  \bibitem{Uhl} 
  K. Uhlenbeck, ``Moduli spaces and adiabatic limits,'' Notices of the AMS, {\bf 42} (1995) 41 .
 
 \bibitem{PS}   
  A.N. Pressley and G.B. Segal, {\it Loop groups}, Oxford University Press, Oxford, 1984.
 
 \bibitem{HS}
  J.A.~Harvey and A.~Strominger,
  ``String theory and the Donaldson polynomial,''\\
  Commun.\ Math.\ Phys.\  {\bf 151} (1993) 221
  [hep-th/9108020].
  
 \bibitem{Gau}
  J.P.~Gauntlett,
  ``Low-energy dynamics of $\Ncal =2$ supersymmetric monopoles,''\\
  Nucl.\ Phys.\ B {\bf 411} (1994) 443
  [hep-th/9305068].
  
 \bibitem{Ale} 
 D.~Alekseevsky and A.~Arvanitoyeorgos, ``Riemannian flag manifolds with homogeneous geodesics,''
 Trans. Amer. Math. Soc. {\bf 359} (2007) 3769.
 
  
\end{thebibliography}
\end{document}